\documentclass[a4paper,australian,amsmath,aps,prl,twocolumn,superscriptaddress,preprintnumbers,noshowpacs]{revtex4-1}
\usepackage{times}
\usepackage{graphicx}
\usepackage{amssymb}
\usepackage{mathrsfs}

\usepackage[usenames]{color}
\definecolor{Green}{rgb}{0.1, 0.8, 0.1}
\definecolor{Orange}{rgb}{1.0, 0.6, 0.4}
\definecolor{Black}{rgb}{0., 0., 0.}
\definecolor{darkgreen}{cmyk}{1,0,1,0.5}
\definecolor{darkblue}{cmyk}{1,1,0,0.1}
\definecolor{r1}{rgb}{.75, .10, .10}

\newcommand{\be}{\begin{equation}}
\newcommand{\ee}{\end{equation}}

\newcommand{\CSSM}{Special Research Centre for the Subatomic Structure
  of Matter (CSSM),\\Department of Physics, University of
  Adelaide, Adelaide, South Australia 5005, Australia} 

\newcommand{\CoEPP}{ARC Centre of Excellence for Particle Physics at
  the Terascale (CoEPP),\\Department of Physics, University
  of Adelaide, Adelaide, South Australia 5005, Australia} 

\newcommand{\NCI}{National Computational Infrastructure
  (NCI),\\Australian National University, Canberra, Australian Capital Territory
  0200, Australia}

\begin{document}

\preprint{ADP-14-34/T893}

\title{Lattice QCD Evidence that the $\mathbf{\Lambda(1405)}$ Resonance is an Antikaon-Nucleon Molecule}

\author{Jonathan M. M. Hall} 
\author{Waseem Kamleh} 
\author{Derek B. Leinweber}
\affiliation{\CSSM}
\author{Benjamin J. Menadue}
\affiliation{\CSSM}
\affiliation{\NCI}
\author{Benjamin J. Owen}
\affiliation{\CSSM}
\author{Anthony W. Thomas} 
\author{Ross D. Young} 
\affiliation{\CSSM}
\affiliation{\CoEPP}


\begin{abstract}
  For almost 50 years the structure of the $\Lambda(1405)$ resonance
  has been a mystery.  Even though it contains a heavy strange quark
  and has odd parity, its mass is lower than any other excited
  spin-1/2 baryon.  Dalitz and co-workers speculated that it might be
  a molecular state of an antikaon bound to a nucleon.  However, a
  standard quark-model structure is also admissible.  Although the
  intervening years have seen considerable effort, there has been no
  convincing resolution.  Here we present a new lattice QCD simulation
  showing that the strange magnetic form factor of the $\Lambda(1405)$
  vanishes, signaling the formation of an antikaon-nucleon molecule.
  Together with a Hamiltonian effective-field-theory model analysis of
  the lattice QCD energy levels, this strongly suggests that the
  structure is dominated by a bound antikaon-nucleon component. This
  result clarifies that not all states occurring in nature can be
  described within a simple quark model framework and points to the
  existence of exotic molecular meson-nucleon bound states.
\end{abstract}

\maketitle

%
The spectrum of hadronic excitations observed at accelerator
facilities around the world manifests the fundamental interactions of
elementary quarks and gluons, governed by the quantum field theory of
quantum chromodynamics (QCD). Understanding the complex emergent
phenomena of this field theory has captivated the attention of
theoretical physicists for more than four decades.

Of particular interest is the unusual nature of the lowest-lying
excitation of the Lambda baryon 
\cite{Alston:1961zzd,Engler:1965zz,Hemingway:1984pz,Riley:1975rg,%
      Esmaili:2009iq,Agakishiev:2012xk,Agashe:2014kda,Hassanvand:2012dn}
the ``Lambda 1405,'' $\Lambda(1405)$.  The Lambda baryon is a neutral
particle, like the neutron, composed of the familiar up $(u)$ and down
$(d)$ quarks together with a strange quark $(s)$.

For almost 50 years the structure of the $\Lambda(1405)$ resonance has
been a mystery.  Even though it contains a relatively massive strange
quark and has odd parity, {\em both of which should increase its
  mass}, it is, in fact, lighter than any other excited spin-1/2
baryon.  Identifying the explanation for this observation has
challenged theorists since its discovery in the 1960s through
kaon-proton \cite{Alston:1961zzd} and pion-proton production
\cite{Engler:1965zz} experiments.

While the quantum numbers of the $\Lambda(1405)$ can be described by
three quarks, $(uds)$, its totally unexpected position in the spectrum
has rendered its structure quite mysterious \cite{Close:1980ab}.
Before the quark model had been established, Dalitz and
co-workers~\cite{Dalitz:1960du,Dalitz:1967fp} speculated that it might
be a molecular state of an antikaon, $\overline{K}$, bound to a
nucleon, $N$.
Whereas the $\pi \Sigma$ energy threshold is well below the
$\Lambda(1405)$ resonance position the $\overline{K}N$ energy
threshold is only slightly above.  A molecular $\overline{K}N$ bound
state with a small amount of binding energy presents an interesting
candidate for the structure of the $\Lambda(1405)$.
Although the intervening years have seen enormous effort
devoted to this resonance 
\cite{Dalitz:1960du,Dalitz:1967fp,Close:1980ab,Veit:1984an,%
  Veit:1984jr,Kaiser:1995eg,Lage:2009zv,Jido:2010ag,%
  Hassanvand:2012dn,Doring:2012eu,Mai:2012dt,Miyagawa:2012xz,%
  MartinezTorres:2012yi,Roca:2013av,Sekihara:2013sma,Xie:2013wfa,%
  Oller:2013zda},
there has been no convincing resolution.

Herein, we present the very first lattice QCD calculation of the
electromagnetic form factors of the $\Lambda(1405)$.  This calculation
reveals the vanishing of the strange quark contribution to the
magnetic form factor of the $\Lambda(1405)$ in the regime where the
masses of the up and down quarks approach their physical values.  This
result is very naturally explained if the state becomes a molecular
$\overline{K}N$ bound state in that limit. When this observation is
combined with a Hamiltonian effective-field-theory analysis of the
structure of the state as a function of its light quark mass, which
shows $\overline{K}N$ dominance and a rapidly decreasing wave function
renormalization constant in the same limit, it constitutes strong
evidence that the $\Lambda(1405)$ is a bound $\overline K N$ molecule.

%
Our calculations are based on the $32^3 \times 64$ full-QCD ensembles
created by the PACS-CS collaboration~\cite{Aoki:2008sm}, made
available through the International Lattice Data Grid (ILDG)
\cite{Beckett:2009cb}. These ensembles provide a lattice volume of
$(2.90\ \mbox{fm})^3$ with five different masses for the light $u$ and
$d$ quarks and constant strange-quark simulation parameters.  We
simulate the valence strange quark with a hopping parameter (governing
the strange quark mass) of $\kappa_s = 0.13665$.  This value
reproduces the correct kaon mass in the physical limit
\cite{Menadue:2012kc}.  We use the squared pion mass as a
renormalization group invariant measure of the quark mass. The
lightest PACS-CS ensemble provides a pion mass of 156 MeV, only
slightly above the physical value of 140 MeV realized in nature.

%
To study the $\Lambda(1405)$ energy, one acts on the QCD vacuum with
baryon interpolating fields defining specific spin, flavor, color
and parity symmetries for the quark field operators.
We consider local three-quark operators providing both scalar and
vector diquark spin configurations for the
quarks~\cite{Menadue:2011pd}.  As the $\Lambda(1405)$ has the valence
quark assignment of $(uds)$, it has overlap with both octet and
singlet flavor symmetries.  The flavor symmetry is exact when all
three quarks have the same mass and electromagnetic charges are
neglected.  However, the strange quark mass is much larger than the
$u$ and $d$ quark masses, and one expects that the eigenstates of QCD
should involve a superposition of octet and singlet symmetries.

The mixing of spin-flavor symmetries in the QCD eigenstates demands a
linear superposition of these interpolators when creating an
eigenstate of QCD.  The correlation matrix
approach~\cite{Michael:1985ne,Luscher:1990ck} provides an effective
means for determining this superposition
via generalized eigenvalue equations.

The basis and superposition of three-quark interpolating fields
successfully isolating the $\Lambda(1405)$ were determined in
Ref.~\cite{Menadue:2011pd} and are used in the present study of the
quark-sector contributions to the electromagnetic form factors of the
$\Lambda(1405)$.
The superposition is illustrated in Fig.~1 of
Ref.~\cite{Menadue:2013xqa} and is characterized by a dominant
flavor-singlet component with the flavor-octet vector-diquark
interpolator gaining importance as the $u$ and $d$ quarks become light
and $SU(3)$ flavor symmetry is broken.  Narrow and wide
gauge-invariant Gaussian-smeared quark sources are considered with the
narrow smearing diminishing in importance as the $u$ and $d$ quarks
become light.

The success of this approach in accurately isolating the
$\Lambda(1405)$, even at the lightest quark mass considered, is
illustrated by the long Euclidean-time single-state stability of the
$\Lambda(1405)$ two-point correlation function presented in Fig.~2 of
Ref.~\cite{Menadue:2013xqa}.  We note this is realized without resort
to two-particle interpolators.  This discovery of a low-lying
$\Lambda(1405)$ mass \cite{Menadue:2011pd} has since been confirmed
independently \cite{Engel:2012qp}.

\begin{figure}[t]
\includegraphics[height=0.98\columnwidth,angle=90]{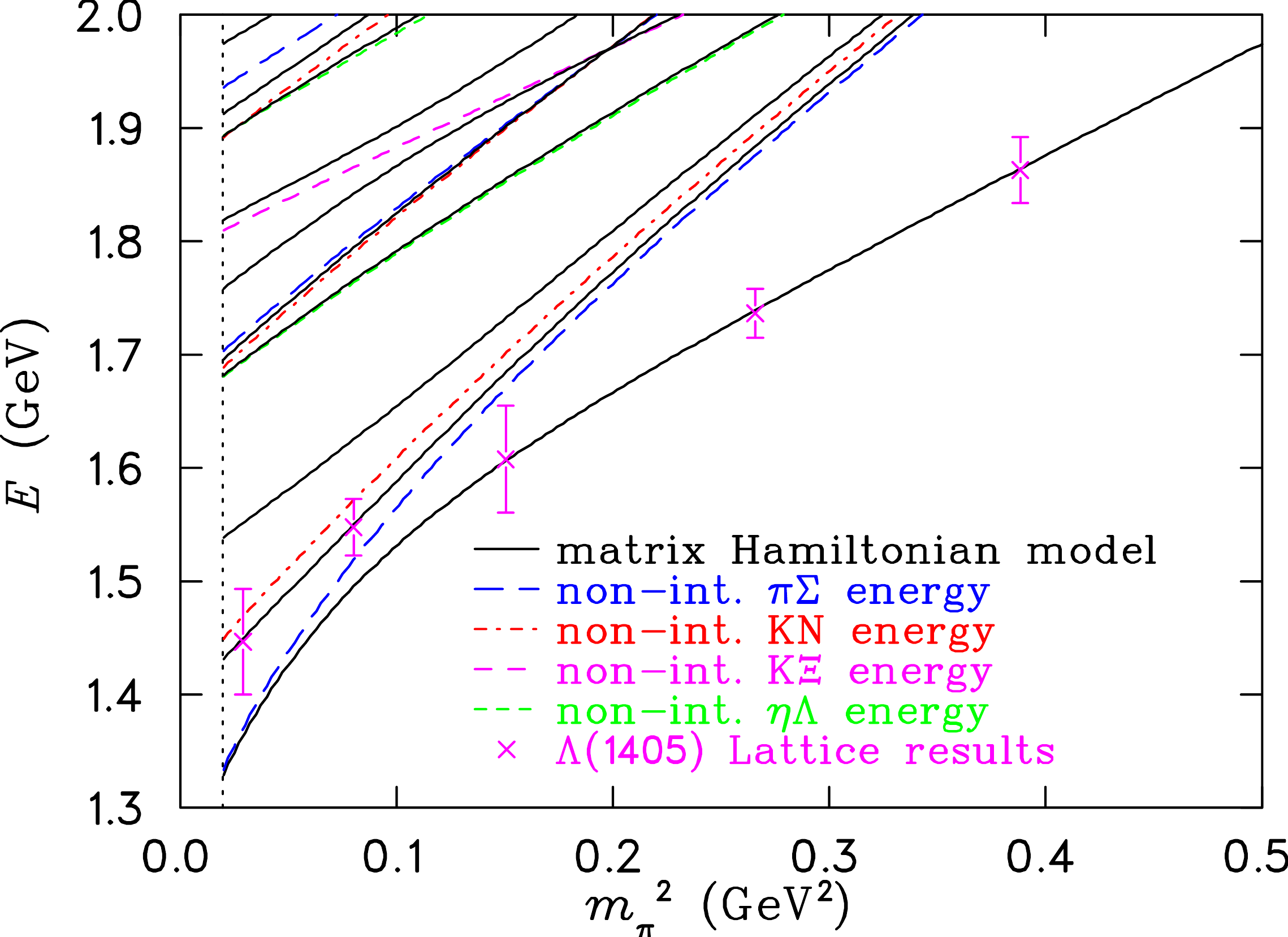}
\vspace*{-8pt}
\caption{The quark-mass dependence $(m_q \propto m_\pi^2)$ of the
  lowest-lying $\Lambda(1405)$ states observed in our lattice QCD
  calculations is illustrated by the discrete points at each of the
  pion masses available in the PACS-CS ensembles.  The low-lying
  energy spectrum of our Hamiltonian model (solid curves) constrained
  to the lattice QCD results (discrete points) is also illustrated.
  The associated noninteracting meson-baryon basis states are
  illustrated by the dashed curves and the
  vertical dashed line indicates the physical pion mass.
\vspace*{-12pt}
\label{fig:modelfin}}
\end{figure}

%
The quark-mass dependence of the lowest-lying state observed in our
lattice QCD calculations is illustrated in Fig.~\ref{fig:modelfin} by
the discrete points at each of the pion masses available in the
PACS-CS simulations.  
The scale is set via the Sommer parameter \cite{Sommer:1993ce} with
$r_0 = 0.492$ fm \cite{Aoki:2008sm}.
This low-lying state is predominantly flavor-singlet,
with an important contribution from octet interpolators emerging as
one moves away from the $SU(3)$ flavor-symmetric limit
\cite{Menadue:2011pd}.

%
The connection of these lattice QCD results obtained on a finite
volume lattice to the infinite volume limit of nature is made
through a matrix Hamiltonian model which describes the composition of
the lattice QCD eigenstates in terms of effective meson-baryon degrees
of freedom.  The noninteracting meson-baryon basis states and the
results of the model are illustrated in Fig.~\ref{fig:modelfin}
and will be discussed in further detail below.

%
The isolation of an individual energy eigenstate enables the
investigation of other properties of the $\Lambda(1405)$ on the finite
volume lattice.  The electromagnetic form factors are particularly
interesting as they provide insight into the distribution of charge
and magnetism within the $\Lambda(1405)$.  Moreover, the form factors
can be resolved one quark flavor at a time.

%
The strange quark magnetic form factor of the $\Lambda(1405)$ is
crucial to the present analysis because it provides direct insight
into the possible dominance of a molecular $\overline{K}N$ bound
state.  In forming such a molecular state, the $\Lambda (uds)$
valence quark configuration is complemented by a $u \overline u$
quark--anti-quark pair making a $K^-(s \overline u)$ proton $(uud)$
bound state, or a $d \overline d$ quark--anti-quark pair making a
${\smash{\overline{K}}\vphantom{K}}^0(s \overline d)$ neutron
$(ddu)$ bound state.  In both cases the strange quark is confined
within a spin-0 kaon and has no preferred spin orientation. Because of
this and the fact that the antikaon has zero orbital angular momentum
in order to conserve parity, the strange quark {\em cannot} contribute
to the magnetic form factor of the $\Lambda(1405)$.  On the other
hand, if the $\Lambda(1405)$ were a $\pi \Sigma$ state or an
elementary three-quark state the strange quark must make a sizable
contribution to the magnetic form factor.  In summary, only if the
$\overline{K}N$ component in the structure of the $\Lambda(1405)$ is
dominant would one expect to find a vanishing strange quark magnetic
form factor.

%
Techniques for calculating the Sachs electromagnetic form factors of
spin-1/2 baryons in lattice QCD were established in
Ref.~\cite{Leinweber:1990dv}.  A fixed boundary condition is applied
in the time direction at $t=0$ and the fermion sources are placed at
$t=16$.  States are evolved in Euclidean time to $t=21$ where an
${\mathcal O}(a)$-improved conserved vector current
\cite{Martinelli:1990ny,Boinepalli:2006xd} is inserted with three
momentum $\vec q = (2\pi/L,0,0)$ providing $Q^2 = 0.16(1)$ GeV${}^2/c^2$.

%
To measure the electromagnetic properties of the $\Lambda(1405)$ in
lattice QCD, one probes the state with the conserved vector current
at a time well separated from the creation and annihilation points to
ensure single-state isolation.  By taking the ratio of this
three-point correlation function with the two-point correlation
function from the mass analysis, we create a direct measure of the
Sachs electric and magnetic form factors.

\begin{figure}[t]
\includegraphics[width=0.98\columnwidth,angle=0]{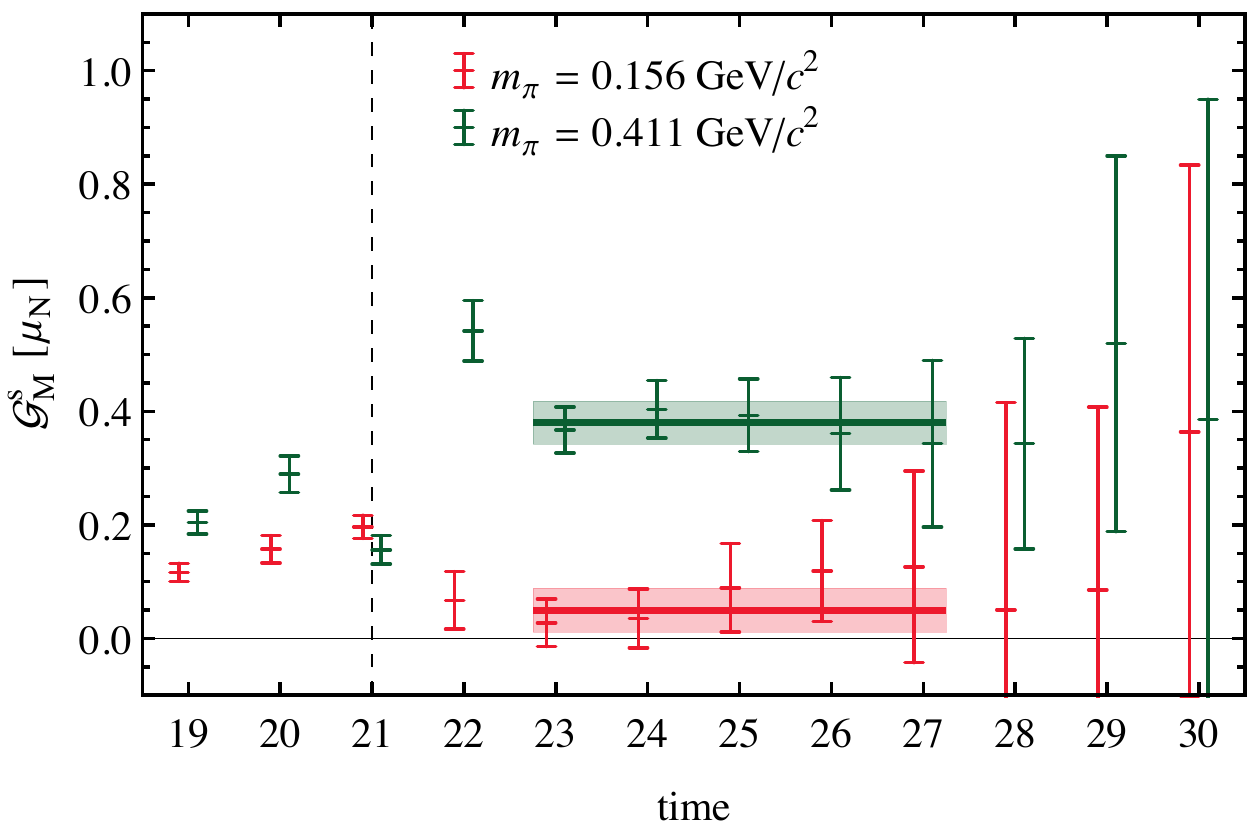}
\vspace*{-8pt}
\caption{Ratio of three- and two-point functions providing the
  strange-quark contribution to the Sachs magnetic form factor of the
  $\Lambda(1405)$.  Results at two different pion masses describing
  the light $u$ and $d$ quark masses are illustrated for $Q^2 \simeq
  0.16$ GeV${}^2/c^2$.  The vertical dashed line indicates the
  introduction of the electromagnetic current at $t=21$ following the
  baryon source at $t=16$.
\vspace*{-12pt}
\label{fig:scomp}}
\end{figure}

%
Figure~\ref{fig:scomp} presents the Euclidean time dependence of this
measure for the strange quark contribution to the Sachs magnetic form
factor, $\mathcal{G}_M^s$, of the $\Lambda(1405)$ at $Q^2 \simeq 0.16$
GeV${}^2/c^2$.  Results for two different ensembles are presented.  As
is standard for quark-sector contributions, the electric charge factor
for the quark charge has not been included; {\it i.e.} the result is
for a single quark of unit charge.  The best fit plateaus, as
identified by a covariance matrix based $\chi^2$ analysis, are also
illustrated.  The rapid onset of the plateau following the
electromagnetic current at $t=21$ reflects our use of optimized
interpolating fields.  

\begin{figure}[t]
\includegraphics[width=0.98\columnwidth]{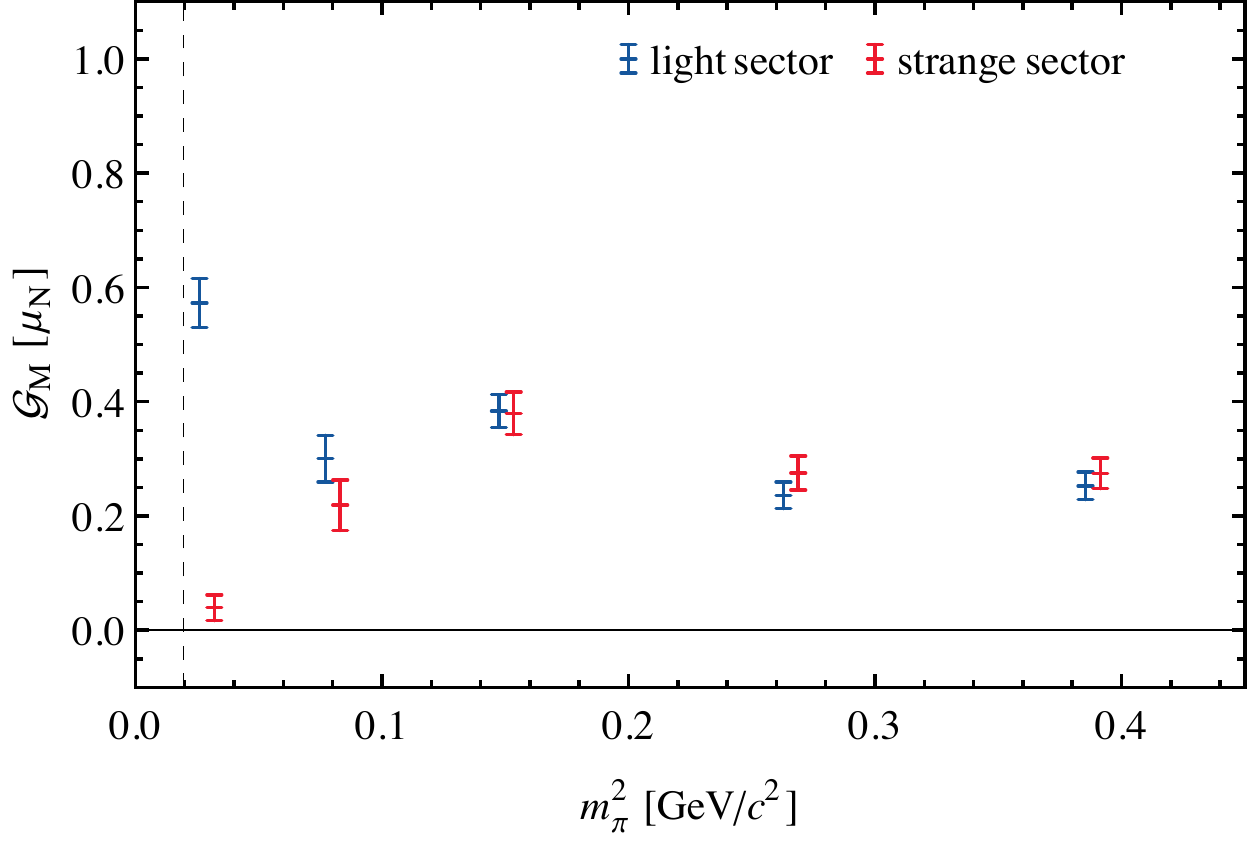}
\vspace*{-8pt}
\caption{The light ($u$ or $d$) and strange ($s$) quark contributions
  to the magnetic form factor of the $\Lambda(1405)$ at $Q^2 \simeq
  0.16$ GeV${}^2/c^2$ are presented as a function of the light $u$ and
  $d$ quark masses, indicated by the squared pion mass, $m_\pi^2$.
  Sector contributions are for single quarks of unit charge.  The
  vertical dashed line indicates the physical pion mass.
\vspace*{-10pt}
\label{fig:GM}}
\end{figure}

%
Figure~\ref{fig:GM} presents ${\mathcal{G}_{M}}^s$ for the
$\Lambda(1405)$ at $Q^2 \simeq 0.16$ GeV${}^2/c^2$ for all five
ensembles available from PACS-CS.  Variation of the light $u$ and $d$
quark masses is indicated by the squared pion mass, $m_\pi^2$.  At the
heaviest $u$ and $d$ quark masses approaching the $SU(3)$ flavor
limit, $m_u = m_d = m_s$, the underlying approximate flavor-singlet
structure is manifest in ${\mathcal{G}_{M}}^s$ with the light and
strange sectors contributing equally.  Similar results were observed
for the electric form factors of the
$\Lambda(1405)$~\cite{Menadue:2013xqa}.  Even though the light-quark
sector is becoming much lighter, this symmetry persists well towards
the physical point.  Only by directly simulating QCD in the realm of
quark masses realized in nature can the vanishing of the strange quark
contribution be revealed.

%
At the lightest quark-mass ensemble closest to nature, the strange
quark contribution to the magnetic form factor of the $\Lambda(1405)$
drops by an order of magnitude and approaches zero.  As the simulation
parameters describing the strange quark are held fixed, this is a
remarkable environmental effect of unprecedented strength.  As the $u$
and $d$ quark masses become light, and the cost of creating $u
\overline u$ and $d \overline d$ quark-antiquark pairs from the QCD
vacuum diminishes, we observe an important rearrangement of the quark
structure within the $\Lambda(1405)$ consistent with the dominance of
a molecular $\overline{K}N$ bound state.

%
To connect these results obtained for a QCD eigenstate on the finite
volume of the lattice to the infinite volume baryon resonance of
nature, we construct a finite-volume Hamiltonian model using a basis
of single- and two-particle noninteracting meson-baryon states
available on the finite-volume periodic lattice.  We follow the
approach established in Ref.~\cite{Hall:2013qba} where the eigenvalue
equation of the model is designed to reproduce finite-volume chiral
effective field
theory~\cite{Savage:1994pf,GarciaRecio:2002td,Sekihara:2010uz,Hyodo:2011ur}
in the weak coupling limit.  Finite-volume
models~\cite{Doring:2012eu,Hall:2013qba,Wu:2014vma} are particularly
useful in interpreting the composition of the energy spectrum observed
in lattice QCD results.

%
In constructing the Hamiltonian, the four octet meson-baryon
interaction channels of the $\Lambda(1405)$ are included: $\pi\Sigma$,
$\overline{K}N$, $K\Xi$ and $\eta\Lambda$.  
The matrix representation of the Hamiltonian contains diagonal entries
corresponding to the relativistic noninteracting meson-baryon
energies available on the finite periodic volume at total
three-momentum zero.  It also includes a single-particle state with a
bare mass parameter, $m_0$.
To access quark masses away from the physical point, the mass of the
bare three-quark state is allowed to grow linearly with the quark
mass (or pion mass squared), $m_0 + \alpha_0\, m_\pi^2$.
The two parameters of the Hamiltonian model, the bare mass, $m_0$, and
the rate of growth, $\alpha_0$, are constrained \cite{Powell} 
by the lattice QCD results.

The interaction entries describe the coupling of the single-particle
state to the two-particle meson-baryon states
\cite{Leinweber:2003dg,Young:2009zb,Beane:2011pc}.
The strength of the interaction is selected to reproduce the physical
decay width (to $\pi \Sigma$) of $50\pm 2$ MeV~\cite{Beringer:1900zz} for the
$\Lambda(1405)$ in the infinite-volume limit.  The couplings for other
channels are related by $SU(3)$-flavor symmetry
\cite{Veit:1984an,Veit:1984jr,Kaiser:1995eg}.

%
In solving the Hamiltonian model, one naturally obtains important
nonperturbative avoided level crossings in the quark mass and volume
dependence of the eigenstates, vital to describing the lattice QCD
results.  The solid curves of Fig.~\ref{fig:modelfin} illustrate the
best fit of the Hamiltonian model to the lattice QCD results.  

%
The three heaviest quark masses considered on the lattice correspond
to a stable odd-parity $\Lambda(1405)$, as the $\pi\Sigma$ threshold
energy exceeds that of the $\Lambda(1405)$.  However, as the physical
pion mass is approached, the $\pi\Sigma$ threshold energy decreases
and a nontrivial mixing of states associated with an avoided level
crossing of the transitioning $\pi\Sigma$ threshold occurs.  At the
lightest two quark masses considered, the $\Lambda(1405)$ corresponds
to the second state of the Hamiltonian model with a
$\pi\Sigma$-dominated eigenstate occupying the lowest energy position.

\begin{figure}[t]
\includegraphics[width=0.98\columnwidth,angle=0]{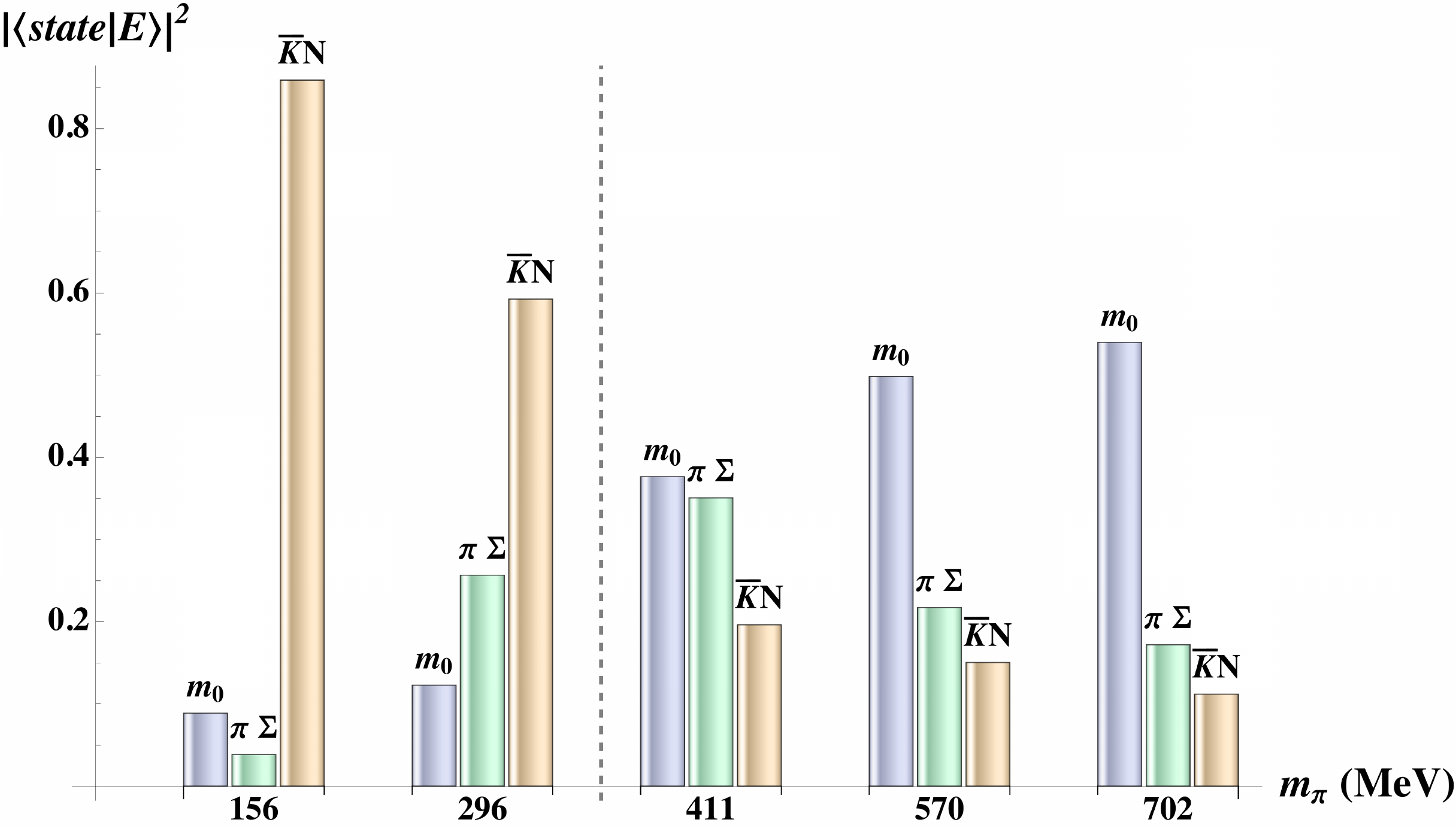}
\vspace*{-8pt}
\caption{The overlap of the basis state, $| {\it state} \rangle$, with
  the energy eigenstate $| E \rangle$ for the $\Lambda(1405)$,
  illustrating the composition of the $\Lambda(1405)$ as a function of
  pion mass.  Basis states include the single particle 
  state, denoted by $m_0,$ and the two-particle states
  $\pi\Sigma$ and $\overline{K}N$.  A sum over all two-particle
  momentum states is done in reporting the probability for the
  two-particle channels.  Pion masses are indicated on the $x$ axis
  with the vertical dashed line separating the first state for the
  heaviest three masses from the second state for the lightest two
  masses.
\vspace*{-10pt}
\label{fig:modelevec}}
\end{figure}

%
The eigenvectors of the Hamiltonian system provide the overlap of the
basis states with the eigenstates and thus describe the underlying
composition of the eigenstates.  As the first and second eigenstates
are dominated by the single-particle state and the two-particle
channels $\pi\Sigma$ and $\overline{K}N$, we illustrate these in
Fig.~\ref{fig:modelevec} for each value of pion mass considered in the
lattice QCD simulations.  A sum over all two-particle momentum states
is done in reporting the probability of the two-particle channels.

%
At the lightest pion mass, $m_\pi = 156$ MeV, the Hamiltonian model
eigenstate for the $\Lambda(1405)$ is dominated by the $\overline{K}N$
channel in complete agreement with the explanation of the observed,
vanishing strange quark contribution to the magnetic form factor.  A
small but nontrivial role for the single-particle three-quark state
enables the excitation of this state in the lattice correlation matrix
analysis of three-quark operators.
%
%
In contrast, the lowest-lying eigenstate of the Hamiltonian model at
$m_\pi = 156$ MeV is dominated by $\pi\Sigma$, with very small
single-particle content, which explains why it is missing from the
lattice QCD spectrum.  

%
Having confirmed that the $\Lambda(1405)$ state observed on the
lattice is best described as a molecular $\overline{K}N$ bound state,
it remains to demonstrate the connection between the finite-volume
lattice eigenstates and the infinite-volume resonance found in nature.
The quark-mass behaviour of the $\Lambda(1405)$ energy in the
infinite-volume limit can be reconstructed from the finite-volume
Hamiltonian model by considering the principal-value continuum
versions of the loop integral contributions from all channels.  A
bootstrap error analysis provides a resonance energy of $1.48{+0.17
  \atop -0.07}$ GeV.
The distribution of the bootstrap analysis is sharply peaked around
the most probable value of 1.41 GeV in good agreement with experiment.
Further details may be found in Ref.~\cite{Hall:2014lat}.

In summary,
the $\Lambda(1405)$ has been identified in first-principles lattice
QCD calculations through a study of its quark mass dependence and its
relation to avoided level crossings in finite-volume effective field
theory.
The structure of the $\Lambda(1405)$ is dominated by a molecular
bound state of an antikaon and a nucleon.
This structure is signified both by the vanishing of the strange quark
contribution to the magnetic moment of the $\Lambda(1405)$ and
by the dominance of the $\overline{K}N$ component found  
in the finite-volume effective field theory Hamiltonian treatment.

At the same time, the presence of a nontrivial single-particle
three-quark component explains why the state is readily accessible in
lattice correlation matrix analyses constructed with three-quark
operators.
In the infinite-volume limit, the Hamiltonian model describes a quark
mass dependence that is consistent with nature.

\begin{acknowledgments}
{\bf Acknowledgements:} We thank PACS-CS Collaboration for making
their gauge configurations available and acknowledge the important
ongoing support of the ILDG.  This research was undertaken with the
assistance of resources awarded at the NCI National Facility in
Canberra, Australia, and the iVEC facilities at Murdoch University
(iVEC@Murdoch) and the University of Western Australia
(iVEC@UWA). These resources are provided through the National
Computational Merit Allocation Scheme and the University of Adelaide
Partner Share supported by the Australian Government.  We also
acknowledge eResearch SA for their support of our supercomputers.
This research is supported by the Australian Research Council through
the ARC Centre of Excellence for Particle Physics at the Terascale,
and through Grants No.\ DP120104627 (D.B.L.), No.\ DP140103067
(D.B.L. and R.D.Y.), No.\ FT120100821 (R.D.Y.) and No.\ FL0992247
(A.W.T.).
\end{acknowledgments}


\begin{thebibliography}{48}%
\makeatletter
\providecommand \@ifxundefined [1]{%
 \@ifx{#1\undefined}
}%
\providecommand \@ifnum [1]{%
 \ifnum #1\expandafter \@firstoftwo
 \else \expandafter \@secondoftwo
 \fi
}%
\providecommand \@ifx [1]{%
 \ifx #1\expandafter \@firstoftwo
 \else \expandafter \@secondoftwo
 \fi
}%
\providecommand \natexlab [1]{#1}%
\providecommand \enquote  [1]{``#1''}%
\providecommand \bibnamefont  [1]{#1}%
\providecommand \bibfnamefont [1]{#1}%
\providecommand \citenamefont [1]{#1}%
\providecommand \href@noop [0]{\@secondoftwo}%
\providecommand \href [0]{\begingroup \@sanitize@url \@href}%
\providecommand \@href[1]{\@@startlink{#1}\@@href}%
\providecommand \@@href[1]{\endgroup#1\@@endlink}%
\providecommand \@sanitize@url [0]{\catcode `\\12\catcode `\$12\catcode
  `\&12\catcode `\#12\catcode `\^12\catcode `\_12\catcode `\%12\relax}%
\providecommand \@@startlink[1]{}%
\providecommand \@@endlink[0]{}%
\providecommand \url  [0]{\begingroup\@sanitize@url \@url }%
\providecommand \@url [1]{\endgroup\@href {#1}{\urlprefix }}%
\providecommand \urlprefix  [0]{URL }%
\providecommand \Eprint [0]{\href }%
\providecommand \doibase [0]{http://dx.doi.org/}%
\providecommand \selectlanguage [0]{\@gobble}%
\providecommand \bibinfo  [0]{\@secondoftwo}%
\providecommand \bibfield  [0]{\@secondoftwo}%
\providecommand \translation [1]{[#1]}%
\providecommand \BibitemOpen [0]{}%
\providecommand \bibitemStop [0]{}%
\providecommand \bibitemNoStop [0]{.\EOS\space}%
\providecommand \EOS [0]{\spacefactor3000\relax}%
\providecommand \BibitemShut  [1]{\csname bibitem#1\endcsname}%
\let\auto@bib@innerbib\@empty
\bibitem [{\citenamefont {Alston}\ \emph {et~al.}(1961)\citenamefont {Alston},
  \citenamefont {Alvarez}, \citenamefont {Eberhard}, \citenamefont {Good},
  \citenamefont {Graziano} \emph {et~al.}}]{Alston:1961zzd}%
  \BibitemOpen
  \bibfield  {author} {\bibinfo {author} {\bibfnamefont {M.~H.}\ \bibnamefont
  {Alston}}, \bibinfo {author} {\bibfnamefont {L.~W.}\ \bibnamefont {Alvarez}},
  \bibinfo {author} {\bibfnamefont {P.}~\bibnamefont {Eberhard}}, \bibinfo
  {author} {\bibfnamefont {M.~L.}\ \bibnamefont {Good}}, \bibinfo {author}
  {\bibfnamefont {W.}~\bibnamefont {Graziano}},  \emph {et~al.},\ }\href
  {\doibase 10.1103/PhysRevLett.6.698} {\bibfield  {journal} {\bibinfo
  {journal} {Phys.Rev.Lett.}\ }\textbf {\bibinfo {volume} {6}},\ \bibinfo
  {pages} {698} (\bibinfo {year} {1961})}\BibitemShut {NoStop}%
\bibitem [{\citenamefont {Engler}\ \emph {et~al.}(1965)\citenamefont {Engler},
  \citenamefont {Fisk}, \citenamefont {Kraemer}, \citenamefont {Meltzer},
  \citenamefont {Westgard} \emph {et~al.}}]{Engler:1965zz}%
  \BibitemOpen
  \bibfield  {author} {\bibinfo {author} {\bibfnamefont {A.}~\bibnamefont
  {Engler}}, \bibinfo {author} {\bibfnamefont {H.}~\bibnamefont {Fisk}},
  \bibinfo {author} {\bibfnamefont {R.}~\bibnamefont {Kraemer}}, \bibinfo
  {author} {\bibfnamefont {C.}~\bibnamefont {Meltzer}}, \bibinfo {author}
  {\bibfnamefont {J.}~\bibnamefont {Westgard}},  \emph {et~al.},\ }\href
  {\doibase 10.1103/PhysRevLett.15.224} {\bibfield  {journal} {\bibinfo
  {journal} {Phys.\ Rev.\ Lett.}\ }\textbf {\bibinfo {volume} {15}},\ \bibinfo
  {pages} {224} (\bibinfo {year} {1965})}\BibitemShut {NoStop}%
\bibitem [{\citenamefont {Hemingway}(1985)}]{Hemingway:1984pz}%
  \BibitemOpen
  \bibfield  {author} {\bibinfo {author} {\bibfnamefont {R.}~\bibnamefont
  {Hemingway}},\ }\href {\doibase 10.1016/0550-3213(85)90556-5} {\bibfield
  {journal} {\bibinfo  {journal} {Nucl.Phys.}\ }\textbf {\bibinfo {volume}
  {B253}},\ \bibinfo {pages} {742} (\bibinfo {year} {1985})}\BibitemShut
  {NoStop}%
\bibitem [{\citenamefont {Riley}\ \emph {et~al.}(1975)\citenamefont {Riley},
  \citenamefont {Wang}, \citenamefont {Fetkovich},\ and\ \citenamefont
  {Mckenzie}}]{Riley:1975rg}%
  \BibitemOpen
  \bibfield  {author} {\bibinfo {author} {\bibfnamefont {B.}~\bibnamefont
  {Riley}}, \bibinfo {author} {\bibfnamefont {I.}~\bibnamefont {Wang}},
  \bibinfo {author} {\bibfnamefont {J.}~\bibnamefont {Fetkovich}}, \ and\
  \bibinfo {author} {\bibfnamefont {J.}~\bibnamefont {Mckenzie}},\ }\href
  {\doibase 10.1103/PhysRevD.11.3065} {\bibfield  {journal} {\bibinfo
  {journal} {Phys.Rev.}\ }\textbf {\bibinfo {volume} {D11}},\ \bibinfo {pages}
  {3065} (\bibinfo {year} {1975})}\BibitemShut {NoStop}%
\bibitem [{\citenamefont {Esmaili}\ \emph {et~al.}(2010)\citenamefont
  {Esmaili}, \citenamefont {Akaishi},\ and\ \citenamefont
  {Yamazaki}}]{Esmaili:2009iq}%
  \BibitemOpen
  \bibfield  {author} {\bibinfo {author} {\bibfnamefont {J.}~\bibnamefont
  {Esmaili}}, \bibinfo {author} {\bibfnamefont {Y.}~\bibnamefont {Akaishi}}, \
  and\ \bibinfo {author} {\bibfnamefont {T.}~\bibnamefont {Yamazaki}},\ }\href
  {\doibase 10.1016/j.physletb.2010.01.075} {\bibfield  {journal} {\bibinfo
  {journal} {Phys.Lett.}\ }\textbf {\bibinfo {volume} {B686}},\ \bibinfo
  {pages} {23} (\bibinfo {year} {2010})},\ \Eprint
  {http://arxiv.org/abs/0906.0505} {arXiv:0906.0505 [nucl-th]} \BibitemShut
  {NoStop}%
\bibitem [{\citenamefont {Agakishiev}\ \emph {et~al.}(2013)\citenamefont
  {Agakishiev} \emph {et~al.}}]{Agakishiev:2012xk}%
  \BibitemOpen
  \bibfield  {author} {\bibinfo {author} {\bibfnamefont {G.}~\bibnamefont
  {Agakishiev}} \emph {et~al.} (\bibinfo {collaboration} {HADES
  Collaboration}),\ }\href {\doibase 10.1103/PhysRevC.87.025201} {\bibfield
  {journal} {\bibinfo  {journal} {Phys.Rev.}\ }\textbf {\bibinfo {volume}
  {C87}},\ \bibinfo {pages} {025201} (\bibinfo {year} {2013})},\ \Eprint
  {http://arxiv.org/abs/1208.0205} {arXiv:1208.0205 [nucl-ex]} \BibitemShut
  {NoStop}%
\bibitem [{\citenamefont {Olive}\ \emph {et~al.}(2014)\citenamefont {Olive}
  \emph {et~al.}}]{Agashe:2014kda}%
  \BibitemOpen
  \bibfield  {author} {\bibinfo {author} {\bibfnamefont {K.}~\bibnamefont
  {Olive}} \emph {et~al.} (\bibinfo {collaboration} {Particle Data Group}),\
  }\href {\doibase 10.1088/1674-1137/38/9/090001} {\bibfield  {journal}
  {\bibinfo  {journal} {Chin.Phys.}\ }\textbf {\bibinfo {volume} {C38}},\
  \bibinfo {pages} {090001} (\bibinfo {year} {2014})}\BibitemShut {NoStop}%
\bibitem [{\citenamefont {Hassanvand}\ \emph {et~al.}(2013)\citenamefont
  {Hassanvand}, \citenamefont {Kalantari}, \citenamefont {Akaishi},\ and\
  \citenamefont {Yamazaki}}]{Hassanvand:2012dn}%
  \BibitemOpen
  \bibfield  {author} {\bibinfo {author} {\bibfnamefont {M.}~\bibnamefont
  {Hassanvand}}, \bibinfo {author} {\bibfnamefont {S.~Z.}\ \bibnamefont
  {Kalantari}}, \bibinfo {author} {\bibfnamefont {Y.}~\bibnamefont {Akaishi}},
  \ and\ \bibinfo {author} {\bibfnamefont {T.}~\bibnamefont {Yamazaki}},\
  }\href {\doibase 10.1103/PhysRevC.87.055202, 10.1103/PhysRevC.88.019905}
  {\bibfield  {journal} {\bibinfo  {journal} {Phys.Rev.}\ }\textbf {\bibinfo
  {volume} {C87}},\ \bibinfo {pages} {055202} (\bibinfo {year} {2013})},\
  \Eprint {http://arxiv.org/abs/1210.7725} {arXiv:1210.7725 [nucl-th]}
  \BibitemShut {NoStop}%
\bibitem [{\citenamefont {Close}\ and\ \citenamefont
  {Dalitz}(1980)}]{Close:1980ab}%
  \BibitemOpen
  \bibfield  {author} {\bibinfo {author} {\bibfnamefont {F.}~\bibnamefont
  {Close}}\ and\ \bibinfo {author} {\bibfnamefont {R.}~\bibnamefont {Dalitz}},\
  }\href@noop {} {\bibfield  {journal} {\bibinfo  {journal} {OXFORD-TP-73-80,
  C80-03-24.2-5}\ } (\bibinfo {year} {1980})}\BibitemShut {NoStop}%
\bibitem [{\citenamefont {Dalitz}\ and\ \citenamefont
  {Tuan}(1960)}]{Dalitz:1960du}%
  \BibitemOpen
  \bibfield  {author} {\bibinfo {author} {\bibfnamefont {R.}~\bibnamefont
  {Dalitz}}\ and\ \bibinfo {author} {\bibfnamefont {S.}~\bibnamefont {Tuan}},\
  }\href@noop {} {\bibfield  {journal} {\bibinfo  {journal} {Annals Phys.}\
  }\textbf {\bibinfo {volume} {10}},\ \bibinfo {pages} {307} (\bibinfo {year}
  {1960})}\BibitemShut {NoStop}%
\bibitem [{\citenamefont {Dalitz}\ \emph {et~al.}(1967)\citenamefont {Dalitz},
  \citenamefont {Wong},\ and\ \citenamefont {Rajasekaran}}]{Dalitz:1967fp}%
  \BibitemOpen
  \bibfield  {author} {\bibinfo {author} {\bibfnamefont {R.}~\bibnamefont
  {Dalitz}}, \bibinfo {author} {\bibfnamefont {T.}~\bibnamefont {Wong}}, \ and\
  \bibinfo {author} {\bibfnamefont {G.}~\bibnamefont {Rajasekaran}},\ }\href
  {\doibase 10.1103/PhysRev.153.1617} {\bibfield  {journal} {\bibinfo
  {journal} {Phys.\ Rev.}\ }\textbf {\bibinfo {volume} {153}},\ \bibinfo
  {pages} {1617} (\bibinfo {year} {1967})}\BibitemShut {NoStop}%
\bibitem [{\citenamefont {Veit}\ \emph {et~al.}(1984)\citenamefont {Veit},
  \citenamefont {Jennings}, \citenamefont {Barrett},\ and\ \citenamefont
  {Thomas}}]{Veit:1984an}%
  \BibitemOpen
  \bibfield  {author} {\bibinfo {author} {\bibfnamefont {E.}~\bibnamefont
  {Veit}}, \bibinfo {author} {\bibfnamefont {B.~K.}\ \bibnamefont {Jennings}},
  \bibinfo {author} {\bibfnamefont {R.}~\bibnamefont {Barrett}}, \ and\
  \bibinfo {author} {\bibfnamefont {A.~W.}\ \bibnamefont {Thomas}},\ }\href
  {\doibase 10.1016/0370-2693(84)91746-5} {\bibfield  {journal} {\bibinfo
  {journal} {Phys.\ Lett.}\ }\textbf {\bibinfo {volume} {B137}},\ \bibinfo
  {pages} {415} (\bibinfo {year} {1984})}\BibitemShut {NoStop}%
\bibitem [{\citenamefont {Veit}\ \emph {et~al.}(1985)\citenamefont {Veit},
  \citenamefont {Jennings}, \citenamefont {Thomas},\ and\ \citenamefont
  {Barrett}}]{Veit:1984jr}%
  \BibitemOpen
  \bibfield  {author} {\bibinfo {author} {\bibfnamefont {E.}~\bibnamefont
  {Veit}}, \bibinfo {author} {\bibfnamefont {B.~K.}\ \bibnamefont {Jennings}},
  \bibinfo {author} {\bibfnamefont {A.~W.}\ \bibnamefont {Thomas}}, \ and\
  \bibinfo {author} {\bibfnamefont {R.}~\bibnamefont {Barrett}},\ }\href
  {\doibase 10.1103/PhysRevD.31.1033} {\bibfield  {journal} {\bibinfo
  {journal} {Phys.\ Rev.}\ }\textbf {\bibinfo {volume} {D31}},\ \bibinfo
  {pages} {1033} (\bibinfo {year} {1985})}\BibitemShut {NoStop}%
\bibitem [{\citenamefont {Kaiser}\ \emph {et~al.}(1995)\citenamefont {Kaiser},
  \citenamefont {Siegel},\ and\ \citenamefont {Weise}}]{Kaiser:1995eg}%
  \BibitemOpen
  \bibfield  {author} {\bibinfo {author} {\bibfnamefont {N.}~\bibnamefont
  {Kaiser}}, \bibinfo {author} {\bibfnamefont {P.}~\bibnamefont {Siegel}}, \
  and\ \bibinfo {author} {\bibfnamefont {W.}~\bibnamefont {Weise}},\ }\href
  {\doibase 10.1016/0375-9474(95)00362-5} {\bibfield  {journal} {\bibinfo
  {journal} {Nucl.\ Phys.}\ }\textbf {\bibinfo {volume} {A594}},\ \bibinfo
  {pages} {325} (\bibinfo {year} {1995})},\ \Eprint
  {http://arxiv.org/abs/nucl-th/9505043} {arXiv:nucl-th/9505043 [nucl-th]}
  \BibitemShut {NoStop}%
\bibitem [{\citenamefont {Lage}\ \emph {et~al.}(2009)\citenamefont {Lage},
  \citenamefont {Meissner},\ and\ \citenamefont {Rusetsky}}]{Lage:2009zv}%
  \BibitemOpen
  \bibfield  {author} {\bibinfo {author} {\bibfnamefont {M.}~\bibnamefont
  {Lage}}, \bibinfo {author} {\bibfnamefont {U.-G.}\ \bibnamefont {Meissner}},
  \ and\ \bibinfo {author} {\bibfnamefont {A.}~\bibnamefont {Rusetsky}},\
  }\href {\doibase 10.1016/j.physletb.2009.10.055} {\bibfield  {journal}
  {\bibinfo  {journal} {Phys.\ Lett.}\ }\textbf {\bibinfo {volume} {B681}},\
  \bibinfo {pages} {439} (\bibinfo {year} {2009})},\ \Eprint
  {http://arxiv.org/abs/0905.0069} {arXiv:0905.0069 [hep-lat]} \BibitemShut
  {NoStop}%
\bibitem [{\citenamefont {Jido}\ \emph {et~al.}(2010)\citenamefont {Jido},
  \citenamefont {Sekihara}, \citenamefont {Ikeda}, \citenamefont {Hyodo},
  \citenamefont {Kanada-En'yo} \emph {et~al.}}]{Jido:2010ag}%
  \BibitemOpen
  \bibfield  {author} {\bibinfo {author} {\bibfnamefont {D.}~\bibnamefont
  {Jido}}, \bibinfo {author} {\bibfnamefont {T.}~\bibnamefont {Sekihara}},
  \bibinfo {author} {\bibfnamefont {Y.}~\bibnamefont {Ikeda}}, \bibinfo
  {author} {\bibfnamefont {T.}~\bibnamefont {Hyodo}}, \bibinfo {author}
  {\bibfnamefont {Y.}~\bibnamefont {Kanada-En'yo}},  \emph {et~al.},\ }\href
  {\doibase 10.1016/j.nuclphysa.2010.01.175} {\bibfield  {journal} {\bibinfo
  {journal} {Nucl.\ Phys.}\ }\textbf {\bibinfo {volume} {A835}},\ \bibinfo
  {pages} {59} (\bibinfo {year} {2010})},\ \Eprint
  {http://arxiv.org/abs/1003.4560} {arXiv:1003.4560 [nucl-th]} \BibitemShut
  {NoStop}%
\bibitem [{\citenamefont {Doring}\ \emph {et~al.}(2012)\citenamefont {Doring},
  \citenamefont {Mei{\ss}ner}, \citenamefont {Oset},\ and\ \citenamefont
  {Rusetsky}}]{Doring:2012eu}%
  \BibitemOpen
  \bibfield  {author} {\bibinfo {author} {\bibfnamefont {M.}~\bibnamefont
  {Doring}}, \bibinfo {author} {\bibfnamefont {U.}~\bibnamefont {Mei{\ss}ner}},
  \bibinfo {author} {\bibfnamefont {E.}~\bibnamefont {Oset}}, \ and\ \bibinfo
  {author} {\bibfnamefont {A.}~\bibnamefont {Rusetsky}},\ }\href {\doibase
  10.1140/epja/i2012-12114-6} {\bibfield  {journal} {\bibinfo  {journal} {Eur.\
  Phys.\ J.}\ }\textbf {\bibinfo {volume} {A48}},\ \bibinfo {pages} {114}
  (\bibinfo {year} {2012})},\ \Eprint {http://arxiv.org/abs/1205.4838}
  {arXiv:1205.4838 [hep-lat]} \BibitemShut {NoStop}%
\bibitem [{\citenamefont {Mai}\ and\ \citenamefont
  {Meissner}(2013)}]{Mai:2012dt}%
  \BibitemOpen
  \bibfield  {author} {\bibinfo {author} {\bibfnamefont {M.}~\bibnamefont
  {Mai}}\ and\ \bibinfo {author} {\bibfnamefont {U.-G.}\ \bibnamefont
  {Meissner}},\ }\href {\doibase 10.1016/j.nuclphysa.2013.01.032} {\bibfield
  {journal} {\bibinfo  {journal} {Nucl.\ Phys.}\ }\textbf {\bibinfo {volume}
  {A900}},\ \bibinfo {pages} {51 } (\bibinfo {year} {2013})},\ \Eprint
  {http://arxiv.org/abs/1202.2030} {arXiv:1202.2030 [nucl-th]} \BibitemShut
  {NoStop}%
\bibitem [{\citenamefont {Miyagawa}\ and\ \citenamefont
  {Haidenbauer}(2012)}]{Miyagawa:2012xz}%
  \BibitemOpen
  \bibfield  {author} {\bibinfo {author} {\bibfnamefont {K.}~\bibnamefont
  {Miyagawa}}\ and\ \bibinfo {author} {\bibfnamefont {J.}~\bibnamefont
  {Haidenbauer}},\ }\href {\doibase 10.1103/PhysRevC.85.065201} {\bibfield
  {journal} {\bibinfo  {journal} {Phys.\ Rev.}\ }\textbf {\bibinfo {volume}
  {C85}},\ \bibinfo {pages} {065201} (\bibinfo {year} {2012})},\ \Eprint
  {http://arxiv.org/abs/1202.4272} {arXiv:1202.4272 [nucl-th]} \BibitemShut
  {NoStop}%
\bibitem [{\citenamefont {Martinez~Torres}\ \emph {et~al.}(2012)\citenamefont
  {Martinez~Torres}, \citenamefont {Bayar}, \citenamefont {Jido},\ and\
  \citenamefont {Oset}}]{MartinezTorres:2012yi}%
  \BibitemOpen
  \bibfield  {author} {\bibinfo {author} {\bibfnamefont {A.}~\bibnamefont
  {Martinez~Torres}}, \bibinfo {author} {\bibfnamefont {M.}~\bibnamefont
  {Bayar}}, \bibinfo {author} {\bibfnamefont {D.}~\bibnamefont {Jido}}, \ and\
  \bibinfo {author} {\bibfnamefont {E.}~\bibnamefont {Oset}},\ }\href {\doibase
  10.1103/PhysRevC.86.055201} {\bibfield  {journal} {\bibinfo  {journal}
  {Phys.\ Rev.}\ }\textbf {\bibinfo {volume} {C86}},\ \bibinfo {pages} {055201}
  (\bibinfo {year} {2012})},\ \Eprint {http://arxiv.org/abs/1202.4297}
  {arXiv:1202.4297 [hep-lat]} \BibitemShut {NoStop}%
\bibitem [{\citenamefont {Roca}\ and\ \citenamefont
  {Oset}(2013)}]{Roca:2013av}%
  \BibitemOpen
  \bibfield  {author} {\bibinfo {author} {\bibfnamefont {L.}~\bibnamefont
  {Roca}}\ and\ \bibinfo {author} {\bibfnamefont {E.}~\bibnamefont {Oset}},\
  }\href {\doibase 10.1103/PhysRevC.87.055201} {\bibfield  {journal} {\bibinfo
  {journal} {Phys.\ Rev.}\ }\textbf {\bibinfo {volume} {C87}},\ \bibinfo
  {pages} {055201} (\bibinfo {year} {2013})},\ \Eprint
  {http://arxiv.org/abs/1301.5741} {arXiv:1301.5741 [nucl-th]} \BibitemShut
  {NoStop}%
\bibitem [{\citenamefont {Sekihara}\ and\ \citenamefont
  {Kumano}(2014)}]{Sekihara:2013sma}%
  \BibitemOpen
  \bibfield  {author} {\bibinfo {author} {\bibfnamefont {T.}~\bibnamefont
  {Sekihara}}\ and\ \bibinfo {author} {\bibfnamefont {S.}~\bibnamefont
  {Kumano}},\ }\href {\doibase 10.1103/PhysRevC.89.025202} {\bibfield
  {journal} {\bibinfo  {journal} {Phys.\ Rev.}\ }\textbf {\bibinfo {volume}
  {C89}},\ \bibinfo {pages} {025202} (\bibinfo {year} {2014})},\ \Eprint
  {http://arxiv.org/abs/1311.4637} {arXiv:1311.4637 [nucl-th]} \BibitemShut
  {NoStop}%
\bibitem [{\citenamefont {Xie}\ \emph {et~al.}(2013)\citenamefont {Xie},
  \citenamefont {Liu},\ and\ \citenamefont {An}}]{Xie:2013wfa}%
  \BibitemOpen
  \bibfield  {author} {\bibinfo {author} {\bibfnamefont {J.-J.}\ \bibnamefont
  {Xie}}, \bibinfo {author} {\bibfnamefont {B.-C.}\ \bibnamefont {Liu}}, \ and\
  \bibinfo {author} {\bibfnamefont {C.-S.}\ \bibnamefont {An}},\ }\href
  {\doibase 10.1103/PhysRevC.88.015203} {\bibfield  {journal} {\bibinfo
  {journal} {Phys.\ Rev.}\ }\textbf {\bibinfo {volume} {C88}},\ \bibinfo
  {pages} {015203} (\bibinfo {year} {2013})},\ \Eprint
  {http://arxiv.org/abs/1307.3707} {arXiv:1307.3707 [nucl-th]} \BibitemShut
  {NoStop}%
\bibitem [{\citenamefont {Oller}(2014)}]{Oller:2013zda}%
  \BibitemOpen
  \bibfield  {author} {\bibinfo {author} {\bibfnamefont {J.}~\bibnamefont
  {Oller}},\ }\href {\doibase 10.1142/S2010194514600969} {\bibfield  {journal}
  {\bibinfo  {journal} {Int.\ J. Mod.\ Phys.\ Conf.\ Ser.}\ }\textbf {\bibinfo
  {volume} {26}},\ \bibinfo {pages} {1460096} (\bibinfo {year} {2014})},\
  \Eprint {http://arxiv.org/abs/1309.2196} {arXiv:1309.2196 [nucl-th]}
  \BibitemShut {NoStop}%
\bibitem [{\citenamefont {Aoki}\ \emph {et~al.}(2009)\citenamefont {Aoki} \emph
  {et~al.}}]{Aoki:2008sm}%
  \BibitemOpen
  \bibfield  {author} {\bibinfo {author} {\bibfnamefont {S.}~\bibnamefont
  {Aoki}} \emph {et~al.} (\bibinfo {collaboration} {PACS-CS Collaboration}),\
  }\href {\doibase 10.1103/PhysRevD.79.034503} {\bibfield  {journal} {\bibinfo
  {journal} {Phys.\ Rev.}\ }\textbf {\bibinfo {volume} {D79}},\ \bibinfo
  {pages} {034503} (\bibinfo {year} {2009})},\ \Eprint
  {http://arxiv.org/abs/0807.1661} {arXiv:0807.1661 [hep-lat]} \BibitemShut
  {NoStop}%
\bibitem [{\citenamefont {Beckett}\ \emph {et~al.}(2011)\citenamefont
  {Beckett}, \citenamefont {Joo}, \citenamefont {Maynard}, \citenamefont
  {Pleiter}, \citenamefont {Tatebe} \emph {et~al.}}]{Beckett:2009cb}%
  \BibitemOpen
  \bibfield  {author} {\bibinfo {author} {\bibfnamefont {M.~G.}\ \bibnamefont
  {Beckett}}, \bibinfo {author} {\bibfnamefont {B.}~\bibnamefont {Joo}},
  \bibinfo {author} {\bibfnamefont {C.~M.}\ \bibnamefont {Maynard}}, \bibinfo
  {author} {\bibfnamefont {D.}~\bibnamefont {Pleiter}}, \bibinfo {author}
  {\bibfnamefont {O.}~\bibnamefont {Tatebe}},  \emph {et~al.},\ }\href
  {\doibase 10.1016/j.cpc.2011.01.027} {\bibfield  {journal} {\bibinfo
  {journal} {Comput.\ Phys.\ Commun.}\ }\textbf {\bibinfo {volume} {182}},\
  \bibinfo {pages} {1208} (\bibinfo {year} {2011})},\ \Eprint
  {http://arxiv.org/abs/0910.1692} {arXiv:0910.1692 [hep-lat]} \BibitemShut
  {NoStop}%
\bibitem [{\citenamefont {Menadue}\ \emph
  {et~al.}(2012{\natexlab{a}})\citenamefont {Menadue}, \citenamefont {Kamleh},
  \citenamefont {Leinweber}, \citenamefont {Mahbub},\ and\ \citenamefont
  {Owen}}]{Menadue:2012kc}%
  \BibitemOpen
  \bibfield  {author} {\bibinfo {author} {\bibfnamefont {B.~J.}\ \bibnamefont
  {Menadue}}, \bibinfo {author} {\bibfnamefont {W.}~\bibnamefont {Kamleh}},
  \bibinfo {author} {\bibfnamefont {D.~B.}\ \bibnamefont {Leinweber}}, \bibinfo
  {author} {\bibfnamefont {M.~S.}\ \bibnamefont {Mahbub}}, \ and\ \bibinfo
  {author} {\bibfnamefont {B.~J.}\ \bibnamefont {Owen}},\ }\href@noop {}
  {\bibfield  {journal} {\bibinfo  {journal} {PoS}\ }\textbf {\bibinfo {volume}
  {LATTICE2012}},\ \bibinfo {pages} {178} (\bibinfo {year}
  {2012}{\natexlab{a}})}\BibitemShut {NoStop}%
\bibitem [{\citenamefont {Menadue}\ \emph
  {et~al.}(2012{\natexlab{b}})\citenamefont {Menadue}, \citenamefont {Kamleh},
  \citenamefont {Leinweber},\ and\ \citenamefont {Mahbub}}]{Menadue:2011pd}%
  \BibitemOpen
  \bibfield  {author} {\bibinfo {author} {\bibfnamefont {B.~J.}\ \bibnamefont
  {Menadue}}, \bibinfo {author} {\bibfnamefont {W.}~\bibnamefont {Kamleh}},
  \bibinfo {author} {\bibfnamefont {D.~B.}\ \bibnamefont {Leinweber}}, \ and\
  \bibinfo {author} {\bibfnamefont {M.~S.}\ \bibnamefont {Mahbub}},\ }\href
  {\doibase 10.1103/PhysRevLett.108.112001} {\bibfield  {journal} {\bibinfo
  {journal} {Phys.\ Rev.\ Lett.}\ }\textbf {\bibinfo {volume} {108}},\ \bibinfo
  {pages} {112001} (\bibinfo {year} {2012}{\natexlab{b}})},\ \Eprint
  {http://arxiv.org/abs/1109.6716} {arXiv:1109.6716 [hep-lat]} \BibitemShut
  {NoStop}%
\bibitem [{\citenamefont {Michael}(1985)}]{Michael:1985ne}%
  \BibitemOpen
  \bibfield  {author} {\bibinfo {author} {\bibfnamefont {C.}~\bibnamefont
  {Michael}},\ }\href {\doibase 10.1016/0550-3213(85)90297-4} {\bibfield
  {journal} {\bibinfo  {journal} {Nucl.\ Phys.}\ }\textbf {\bibinfo {volume}
  {B259}},\ \bibinfo {pages} {58} (\bibinfo {year} {1985})}\BibitemShut
  {NoStop}%
\bibitem [{\citenamefont {Luscher}\ and\ \citenamefont
  {Wolff}(1990)}]{Luscher:1990ck}%
  \BibitemOpen
  \bibfield  {author} {\bibinfo {author} {\bibfnamefont {M.}~\bibnamefont
  {Luscher}}\ and\ \bibinfo {author} {\bibfnamefont {U.}~\bibnamefont
  {Wolff}},\ }\href {\doibase 10.1016/0550-3213(90)90540-T} {\bibfield
  {journal} {\bibinfo  {journal} {Nucl.\ Phys.}\ }\textbf {\bibinfo {volume}
  {B339}},\ \bibinfo {pages} {222} (\bibinfo {year} {1990})}\BibitemShut
  {NoStop}%
\bibitem [{\citenamefont {Menadue}\ \emph {et~al.}(2013)\citenamefont
  {Menadue}, \citenamefont {Kamleh}, \citenamefont {Leinweber}, \citenamefont
  {Mahbub},\ and\ \citenamefont {Owen}}]{Menadue:2013xqa}%
  \BibitemOpen
  \bibfield  {author} {\bibinfo {author} {\bibfnamefont {B.~J.}\ \bibnamefont
  {Menadue}}, \bibinfo {author} {\bibfnamefont {W.}~\bibnamefont {Kamleh}},
  \bibinfo {author} {\bibfnamefont {D.~B.}\ \bibnamefont {Leinweber}}, \bibinfo
  {author} {\bibfnamefont {M.~S.}\ \bibnamefont {Mahbub}}, \ and\ \bibinfo
  {author} {\bibfnamefont {B.~J.}\ \bibnamefont {Owen}},\ }\href@noop {}
  {\bibfield  {journal} {\bibinfo  {journal} {PoS}\ }\textbf {\bibinfo {volume}
  {LATTICE2013}},\ \bibinfo {pages} {280} (\bibinfo {year} {2013})},\ \Eprint
  {http://arxiv.org/abs/1311.5026} {arXiv:1311.5026 [hep-lat]} \BibitemShut
  {NoStop}%
\bibitem [{\citenamefont {Engel}\ \emph {et~al.}(2013)\citenamefont {Engel},
  \citenamefont {Lang},\ and\ \citenamefont {Schafer}}]{Engel:2012qp}%
  \BibitemOpen
  \bibfield  {author} {\bibinfo {author} {\bibfnamefont {G.~P.}\ \bibnamefont
  {Engel}}, \bibinfo {author} {\bibfnamefont {C.}~\bibnamefont {Lang}}, \ and\
  \bibinfo {author} {\bibfnamefont {A.}~\bibnamefont {Schafer}},\ }\href
  {\doibase 10.1103/PhysRevD.87.034502} {\bibfield  {journal} {\bibinfo
  {journal} {Phys.\ Rev.}\ }\textbf {\bibinfo {volume} {D87}},\ \bibinfo
  {pages} {034502} (\bibinfo {year} {2013})},\ \Eprint
  {http://arxiv.org/abs/1212.2032} {arXiv:1212.2032 [hep-lat]} \BibitemShut
  {NoStop}%
\bibitem [{\citenamefont {Sommer}(1994)}]{Sommer:1993ce}%
  \BibitemOpen
  \bibfield  {author} {\bibinfo {author} {\bibfnamefont {R.}~\bibnamefont
  {Sommer}},\ }\href {\doibase 10.1016/0550-3213(94)90473-1} {\bibfield
  {journal} {\bibinfo  {journal} {Nucl.\ Phys.}\ }\textbf {\bibinfo {volume}
  {B411}},\ \bibinfo {pages} {839} (\bibinfo {year} {1994})},\ \Eprint
  {http://arxiv.org/abs/hep-lat/9310022} {arXiv:hep-lat/9310022 [hep-lat]}
  \BibitemShut {NoStop}%
\bibitem [{\citenamefont {Leinweber}\ \emph {et~al.}(1991)\citenamefont
  {Leinweber}, \citenamefont {Woloshyn},\ and\ \citenamefont
  {Draper}}]{Leinweber:1990dv}%
  \BibitemOpen
  \bibfield  {author} {\bibinfo {author} {\bibfnamefont {D.~B.}\ \bibnamefont
  {Leinweber}}, \bibinfo {author} {\bibfnamefont {R.}~\bibnamefont {Woloshyn}},
  \ and\ \bibinfo {author} {\bibfnamefont {T.}~\bibnamefont {Draper}},\ }\href
  {\doibase 10.1103/PhysRevD.43.1659} {\bibfield  {journal} {\bibinfo
  {journal} {Phys.\ Rev.}\ }\textbf {\bibinfo {volume} {D43}},\ \bibinfo
  {pages} {1659} (\bibinfo {year} {1991})}\BibitemShut {NoStop}%
\bibitem [{\citenamefont {Martinelli}\ \emph {et~al.}(1991)\citenamefont
  {Martinelli}, \citenamefont {Sachrajda},\ and\ \citenamefont
  {Vladikas}}]{Martinelli:1990ny}%
  \BibitemOpen
  \bibfield  {author} {\bibinfo {author} {\bibfnamefont {G.}~\bibnamefont
  {Martinelli}}, \bibinfo {author} {\bibfnamefont {C.~T.}\ \bibnamefont
  {Sachrajda}}, \ and\ \bibinfo {author} {\bibfnamefont {A.}~\bibnamefont
  {Vladikas}},\ }\href {\doibase 10.1016/0550-3213(91)90538-9} {\bibfield
  {journal} {\bibinfo  {journal} {Nucl.\ Phys.}\ }\textbf {\bibinfo {volume}
  {B358}},\ \bibinfo {pages} {212} (\bibinfo {year} {1991})}\BibitemShut
  {NoStop}%
\bibitem [{\citenamefont {Boinepalli}\ \emph {et~al.}(2006)\citenamefont
  {Boinepalli}, \citenamefont {Leinweber}, \citenamefont {Williams},
  \citenamefont {Zanotti},\ and\ \citenamefont {Zhang}}]{Boinepalli:2006xd}%
  \BibitemOpen
  \bibfield  {author} {\bibinfo {author} {\bibfnamefont {S.}~\bibnamefont
  {Boinepalli}}, \bibinfo {author} {\bibfnamefont {D.}~\bibnamefont
  {Leinweber}}, \bibinfo {author} {\bibfnamefont {A.}~\bibnamefont {Williams}},
  \bibinfo {author} {\bibfnamefont {J.}~\bibnamefont {Zanotti}}, \ and\
  \bibinfo {author} {\bibfnamefont {J.}~\bibnamefont {Zhang}},\ }\href
  {\doibase 10.1103/PhysRevD.74.093005} {\bibfield  {journal} {\bibinfo
  {journal} {Phys.\ Rev.}\ }\textbf {\bibinfo {volume} {D74}},\ \bibinfo
  {pages} {093005} (\bibinfo {year} {2006})},\ \Eprint
  {http://arxiv.org/abs/hep-lat/0604022} {arXiv:hep-lat/0604022 [hep-lat]}
  \BibitemShut {NoStop}%
\bibitem [{\citenamefont {Hall}\ \emph {et~al.}(2013)\citenamefont {Hall},
  \citenamefont {Hsu}, \citenamefont {Leinweber}, \citenamefont {Thomas},\ and\
  \citenamefont {Young}}]{Hall:2013qba}%
  \BibitemOpen
  \bibfield  {author} {\bibinfo {author} {\bibfnamefont {J.}~\bibnamefont
  {Hall}}, \bibinfo {author} {\bibfnamefont {A.~C.~P.}\ \bibnamefont {Hsu}},
  \bibinfo {author} {\bibfnamefont {D.}~\bibnamefont {Leinweber}}, \bibinfo
  {author} {\bibfnamefont {A.}~\bibnamefont {Thomas}}, \ and\ \bibinfo {author}
  {\bibfnamefont {R.}~\bibnamefont {Young}},\ }\href {\doibase
  10.1103/PhysRevD.87.094510} {\bibfield  {journal} {\bibinfo  {journal}
  {Phys.\ Rev.}\ }\textbf {\bibinfo {volume} {D87}},\ \bibinfo {pages} {094510}
  (\bibinfo {year} {2013})},\ \Eprint {http://arxiv.org/abs/1303.4157}
  {arXiv:1303.4157 [hep-lat]} \BibitemShut {NoStop}%
\bibitem [{\citenamefont {Savage}(1994)}]{Savage:1994pf}%
  \BibitemOpen
  \bibfield  {author} {\bibinfo {author} {\bibfnamefont {M.~J.}\ \bibnamefont
  {Savage}},\ }\href {\doibase 10.1016/0370-2693(94)91072-3} {\bibfield
  {journal} {\bibinfo  {journal} {Phys.\ Lett.}\ }\textbf {\bibinfo {volume}
  {B331}},\ \bibinfo {pages} {411} (\bibinfo {year} {1994})},\ \Eprint
  {http://arxiv.org/abs/hep-ph/9404285} {arXiv:hep-ph/9404285 [hep-ph]}
  \BibitemShut {NoStop}%
\bibitem [{\citenamefont {Garcia-Recio}\ \emph {et~al.}(2003)\citenamefont
  {Garcia-Recio}, \citenamefont {Nieves}, \citenamefont {Ruiz~Arriola},\ and\
  \citenamefont {Vicente~Vacas}}]{GarciaRecio:2002td}%
  \BibitemOpen
  \bibfield  {author} {\bibinfo {author} {\bibfnamefont {C.}~\bibnamefont
  {Garcia-Recio}}, \bibinfo {author} {\bibfnamefont {J.}~\bibnamefont
  {Nieves}}, \bibinfo {author} {\bibfnamefont {E.}~\bibnamefont
  {Ruiz~Arriola}}, \ and\ \bibinfo {author} {\bibfnamefont {M.}~\bibnamefont
  {Vicente~Vacas}},\ }\href {\doibase 10.1103/PhysRevD.67.076009} {\bibfield
  {journal} {\bibinfo  {journal} {Phys.\ Rev.}\ }\textbf {\bibinfo {volume}
  {D67}},\ \bibinfo {pages} {076009} (\bibinfo {year} {2003})},\ \Eprint
  {http://arxiv.org/abs/hep-ph/0210311} {arXiv:hep-ph/0210311 [hep-ph]}
  \BibitemShut {NoStop}%
\bibitem [{\citenamefont {Sekihara}\ \emph {et~al.}(2011)\citenamefont
  {Sekihara}, \citenamefont {Hyodo},\ and\ \citenamefont
  {Jido}}]{Sekihara:2010uz}%
  \BibitemOpen
  \bibfield  {author} {\bibinfo {author} {\bibfnamefont {T.}~\bibnamefont
  {Sekihara}}, \bibinfo {author} {\bibfnamefont {T.}~\bibnamefont {Hyodo}}, \
  and\ \bibinfo {author} {\bibfnamefont {D.}~\bibnamefont {Jido}},\ }\href
  {\doibase 10.1103/PhysRevC.83.055202} {\bibfield  {journal} {\bibinfo
  {journal} {Phys.\ Rev.}\ }\textbf {\bibinfo {volume} {C83}},\ \bibinfo
  {pages} {055202} (\bibinfo {year} {2011})},\ \Eprint
  {http://arxiv.org/abs/1012.3232} {arXiv:1012.3232 [nucl-th]} \BibitemShut
  {NoStop}%
\bibitem [{\citenamefont {Hyodo}\ and\ \citenamefont
  {Jido}(2012)}]{Hyodo:2011ur}%
  \BibitemOpen
  \bibfield  {author} {\bibinfo {author} {\bibfnamefont {T.}~\bibnamefont
  {Hyodo}}\ and\ \bibinfo {author} {\bibfnamefont {D.}~\bibnamefont {Jido}},\
  }\href {\doibase 10.1016/j.ppnp.2011.07.002} {\bibfield  {journal} {\bibinfo
  {journal} {Prog.\ Part.\ Nucl.\ Phys.}\ }\textbf {\bibinfo {volume} {67}},\
  \bibinfo {pages} {55} (\bibinfo {year} {2012})},\ \Eprint
  {http://arxiv.org/abs/1104.4474} {arXiv:1104.4474 [nucl-th]} \BibitemShut
  {NoStop}%
\bibitem [{\citenamefont {Wu}\ \emph {et~al.}(2014)\citenamefont {Wu},
  \citenamefont {Lee}, \citenamefont {Thomas},\ and\ \citenamefont
  {Young}}]{Wu:2014vma}%
  \BibitemOpen
  \bibfield  {author} {\bibinfo {author} {\bibfnamefont {J.-J.}\ \bibnamefont
  {Wu}}, \bibinfo {author} {\bibfnamefont {T.-S.}\ \bibnamefont {Lee}},
  \bibinfo {author} {\bibfnamefont {A.}~\bibnamefont {Thomas}}, \ and\ \bibinfo
  {author} {\bibfnamefont {R.}~\bibnamefont {Young}},\ }\href {\doibase
  10.1103/PhysRevC.90.055206} {\bibfield  {journal} {\bibinfo  {journal}
  {Phys.Rev.}\ }\textbf {\bibinfo {volume} {C90}},\ \bibinfo {pages} {055206}
  (\bibinfo {year} {2014})},\ \Eprint {http://arxiv.org/abs/1402.4868}
  {arXiv:1402.4868 [hep-lat]} \BibitemShut {NoStop}%
\bibitem [{\citenamefont {Powell}(1964)}]{Powell}%
  \BibitemOpen
  \bibfield  {author} {\bibinfo {author} {\bibfnamefont {M.}~\bibnamefont
  {Powell}},\ }\href {\doibase 10.1093/comjnl/7.2.155} {\bibfield  {journal}
  {\bibinfo  {journal} {The Computer Journal}\ }\textbf {\bibinfo {volume}
  {7}},\ \bibinfo {pages} {155} (\bibinfo {year} {1964})}\BibitemShut {NoStop}%
\bibitem [{\citenamefont {Leinweber}\ \emph {et~al.}(2004)\citenamefont
  {Leinweber}, \citenamefont {Thomas},\ and\ \citenamefont
  {Young}}]{Leinweber:2003dg}%
  \BibitemOpen
  \bibfield  {author} {\bibinfo {author} {\bibfnamefont {D.~B.}\ \bibnamefont
  {Leinweber}}, \bibinfo {author} {\bibfnamefont {A.~W.}\ \bibnamefont
  {Thomas}}, \ and\ \bibinfo {author} {\bibfnamefont {R.~D.}\ \bibnamefont
  {Young}},\ }\href {\doibase 10.1103/PhysRevLett.92.242002} {\bibfield
  {journal} {\bibinfo  {journal} {Phys.\ Rev.\ Lett.}\ }\textbf {\bibinfo
  {volume} {92}},\ \bibinfo {pages} {242002} (\bibinfo {year} {2004})},\
  \Eprint {http://arxiv.org/abs/hep-lat/0302020} {arXiv:hep-lat/0302020
  [hep-lat]} \BibitemShut {NoStop}%
\bibitem [{\citenamefont {Young}\ and\ \citenamefont
  {Thomas}(2010)}]{Young:2009zb}%
  \BibitemOpen
  \bibfield  {author} {\bibinfo {author} {\bibfnamefont {R.}~\bibnamefont
  {Young}}\ and\ \bibinfo {author} {\bibfnamefont {A.}~\bibnamefont {Thomas}},\
  }\href {\doibase 10.1103/PhysRevD.81.014503} {\bibfield  {journal} {\bibinfo
  {journal} {Phys.\ Rev.}\ }\textbf {\bibinfo {volume} {D81}},\ \bibinfo
  {pages} {014503} (\bibinfo {year} {2010})},\ \Eprint
  {http://arxiv.org/abs/0901.3310} {arXiv:0901.3310 [hep-lat]} \BibitemShut
  {NoStop}%
\bibitem [{\citenamefont {Beane}\ \emph {et~al.}(2011)\citenamefont {Beane},
  \citenamefont {Chang}, \citenamefont {Detmold}, \citenamefont {Lin},
  \citenamefont {Luu} \emph {et~al.}}]{Beane:2011pc}%
  \BibitemOpen
  \bibfield  {author} {\bibinfo {author} {\bibfnamefont {S.}~\bibnamefont
  {Beane}}, \bibinfo {author} {\bibfnamefont {E.}~\bibnamefont {Chang}},
  \bibinfo {author} {\bibfnamefont {W.}~\bibnamefont {Detmold}}, \bibinfo
  {author} {\bibfnamefont {H.}~\bibnamefont {Lin}}, \bibinfo {author}
  {\bibfnamefont {T.}~\bibnamefont {Luu}},  \emph {et~al.},\ }\href {\doibase
  10.1103/PhysRevD.84.014507} {\bibfield  {journal} {\bibinfo  {journal}
  {Phys.\ Rev.}\ }\textbf {\bibinfo {volume} {D84}},\ \bibinfo {pages} {014507}
  (\bibinfo {year} {2011})},\ \Eprint {http://arxiv.org/abs/1104.4101}
  {arXiv:1104.4101 [hep-lat]} \BibitemShut {NoStop}%
\bibitem [{\citenamefont {Beringer}\ \emph {et~al.}(2012)\citenamefont
  {Beringer} \emph {et~al.}}]{Beringer:1900zz}%
  \BibitemOpen
  \bibfield  {author} {\bibinfo {author} {\bibfnamefont {J.}~\bibnamefont
  {Beringer}} \emph {et~al.} (\bibinfo {collaboration} {Particle Data Group}),\
  }\href {\doibase 10.1103/PhysRevD.86.010001} {\bibfield  {journal} {\bibinfo
  {journal} {Phys.\ Rev.}\ }\textbf {\bibinfo {volume} {D86}},\ \bibinfo
  {pages} {010001} (\bibinfo {year} {2012})}\BibitemShut {NoStop}%
\bibitem [{\citenamefont {Hall}\ \emph {et~al.}(2014)\citenamefont {Hall},
  \citenamefont {Kamleh}, \citenamefont {Leinweber}, \citenamefont {Menadue},
  \citenamefont {Owen}, \citenamefont {Thomas},\ and\ \citenamefont
  {Young}}]{Hall:2014lat}%
  \BibitemOpen
  \bibfield  {author} {\bibinfo {author} {\bibfnamefont {J.~M.~M.}\
  \bibnamefont {Hall}}, \bibinfo {author} {\bibfnamefont {W.}~\bibnamefont
  {Kamleh}}, \bibinfo {author} {\bibfnamefont {D.~B.}\ \bibnamefont
  {Leinweber}}, \bibinfo {author} {\bibfnamefont {B.~J.}\ \bibnamefont
  {Menadue}}, \bibinfo {author} {\bibfnamefont {B.~J.}\ \bibnamefont {Owen}},
  \bibinfo {author} {\bibfnamefont {A.~W.}\ \bibnamefont {Thomas}}, \ and\
  \bibinfo {author} {\bibfnamefont {R.~D.}\ \bibnamefont {Young}},\ }\href@noop
  {} {\bibfield  {journal} {\bibinfo  {journal} {PoS}\ }\textbf {\bibinfo
  {volume} {LATTICE2014}},\ \bibinfo {pages} {094} (\bibinfo {year} {2014})},\
  \Eprint {http://arxiv.org/abs/1411.3781} {arXiv:1411.3781 [hep-lat]}
  \BibitemShut {NoStop}%
\end{thebibliography}

%

\end{document}